\documentclass[preprint]{elsarticle}
\journal{NIM Section A}

\usepackage{graphicx}
\usepackage{dcolumn}
\usepackage{multirow}

\usepackage{bm}
\usepackage[bookmarks]{hyperref}
\usepackage{subfigure}
\usepackage{graphicx}
\usepackage{lscape}
\usepackage{epsfig}
\usepackage{rotating}

\usepackage{epstopdf}
\usepackage{tabularx}
\usepackage{array}

\usepackage{amsmath}
\usepackage{mathtools}
\usepackage{xspace}

\hypersetup{
    colorlinks=true,
    linkcolor=blue,
    filecolor=magenta,      
    urlcolor=cyan,
}

\newcommand\geniso[2]{\ensuremath{\rm ^{#2}#1}\xspace}

\newcommand\newiso[3][]{
\ifx&#1&
  \expandafter\newcommand\csname #2\endcsname[1][#3]{\geniso{#2}{##1}}
\else
  \expandafter\newcommand\csname #1\endcsname[1][#3]{\geniso{#2}{##1}}
\fi
}

\newiso{K}{40}
\newiso{U}{238}
\newiso{Th}{232}
\newiso{Co}{60}
\newiso{Ra}{}
\newiso{Rn}{}
\newiso{Bi}{}
\newiso{Pb}{210}
\newiso{Tl}{208}
\newiso{Ac}{}
\newiso{W}{}
\newiso{Hf}{}
\newiso{Cs}{}
\newiso{Ca}{}
\newiso{Sb}{}
\newiso{La}{}
\newiso{Cr}{}
\newiso{Zn}{}
\newiso{As}{}
\newiso{Au}{}
\newiso{Ti}{}
\newiso{Sc}{}
\newiso{Xe}{}
\newiso{Ar}{40}
\newiso{Na}{22}
\newiso{I}{125}
\newiso{Te}{}
\newiso{Po}{}
\newiso{Am}{241}
\newiso{Fe}{55}
\newiso{Mo}{100}
\newiso{Zr}{}
\newiso{Nd}{}
\newiso{Tc}{}
\newiso{Ba}{}
\newiso{Y}{}
\newiso{Ta}{}

\begin{document}
\begin{frontmatter}

\title{Development of an array of HPGe detectors with 980\% relative efficiency}

\author[CUP]{D.~S. Leonard}
\ead{dleonard@ibs.re.kr}

\author[EWHAEd,CENS]{I.~S.~Hahn}
\author[CUP]{W.~G.~Kang}
\author[Baksan]{V.~Kazalov}
\author[CUP]{G.~W.~Kim}
\author[CUP,Sejong,UST]{Y.~D.~Kim}
\author[CUP]{E.~K.~Lee}
\author[CUP,UST]{M.~H.~Lee\corref{cor}}
\cortext[cor]{Corresponding author.}
\ead{mhlee @ ibs.re.kr}

\author[EWHAPhys]{S.~Y.~Park}

\author[CUP]{E.~Sala\corref{cor2}\fnref{Sala}}
\cortext[cor2]{Corresponding author.}
\ead{elena.sala@ibs.re.kr}
\fntext[Sala]{Now at Center for Axion and Precision Physics Research, Institute for Basic Science, Daejeon 34051, Korea}

\address[CUP]{Center for Underground Physics, Institute for Basic Science, Daejeon 34126, Korea}
\address[EWHAEd]{Department of Science Education, Ewha Womans University, Seoul 03760, Korea}
\address[CENS]{Center for Exotic Nuclear Studies, Institute for Basic Science, Daejeon 34126, Korea}
\address[Baksan]{Baksan Neutrino Observatory, Institute for Nuclear Research of the Russian Academy of Science, Kabardino-Balkaria 361609, Russia}
\address[Sejong]{Department of Physics and Astronomy, Sejong University, Seoul 05006, Korea}

\address[UST]{IBS School, University of Science and Technology, Daejeon, 34113, Korea}
\address[EWHAPhys]{Department of Physics, Ewha Womans University, Seoul 03760, Korea}

\begin{abstract}
Searches for new physics push experiments to look for increasingly rare interactions.  As a result, detectors require increasing sensitivity and specificity, and materials must be screened for naturally occurring, background-producing radioactivity.  Furthermore the detectors used for screening must approach the sensitivities of the physics-search detectors themselves, thus motivating iterative development of detectors capable of both physics searches and background screening.  We report on the design, installation, and performance of a novel, low-background, fourteen-element high-purity germanium detector named the CAGe (CUP Array of Germanium), installed at the Yangyang underground laboratory in Korea. 
\end{abstract}

\end{frontmatter}


\section{Introduction}
\label{sec:Intro}

The Center for Underground Physics (CUP) at the Institute for Basic Science (IBS) and collaborating institutions are involved in developing and operating a number of rare-event physics-search experiments, including searches for dark matter and neutrinoless double-beta decay.   Probably the largest challenge to all such experiments is controlling and understanding the experiment background signals arising from trace levels of naturally occurring radioactive contaminants in the detector construction materials. In collaboration with CANBERRA (now MIRION Technologies), we have customized and installed a fourteen-element high-purity germanium (HPGe) array detector named the CAGe (CUP Array of Germanium) that is intended for screening detector materials for trace levels of radioactivity~\cite{Mo100_ICHEP,Mo100_JKPS,HPGe_ICHEP,Assay_TAUP} and for directly performing new physics searches such as searches for the decay of \Ta[180m]~\cite{Ta_sim,Ta180_PhD}. 

Because of their large penetration depths, gamma emissions in particular, including indirect bremsstrahlung photons from beta-decays, place contamination constraints on all materials in and around a low-rate detector, making these backgrounds one of the most general classes of concerns for such experiments.  In particular, backgrounds arise from gamma-emitting decays within decay chains supported by long-lived naturally occurring radioactive isotopes such as \K[40], \Th[228], and \Ra[226], as well cosmologically activated isotopes such as \Co[60].

The background rates detectable and relevant to development of the most sensitive detectors in the world tend to be lower, by design, than the sensitivity levels of existing detectors, making it very challenging to screen materials for construction of the new detectors. For this reason, the screening detectors must be developed along with the physics detectors, and can also then serve both purposes. 

An optimal gamma-screening detector should have a high efficiency and strong discrimination of signals from background, also qualities of a good physics-search detector.  By using an array of fourteen HPGe detectors with 70\% relative efficiency each, we can obtain a large overall efficiency. The segmented nature of the arrangement also allows for detection of coincident signals, which when correlated to gamma energy can often be used to enhance signal detection or background rejection.  In particular, signals arising from internal samples are, in comparison to those from external shielding elements, relatively more likely to produce multiple-detector, multiple-gamma coincidence events, identifiable in cases where at least one of the multiple detector signals corresponds to a full-energy gamma peak. We note similarities to the single-crysotat array reported in Ref.~\cite{CASCADES}, which itself was inspired by the even more similar fourteen-element conceptual design presented in Ref.~\cite{CASCADES_concept}.

In order to optimize the utility of the array, we carried out an extensive campaign to characterize a prototype single-element detector and to screen and customize array components to reduce the final array backgrounds. The array was installed in the A5 tunnel at the Yangyang underground laboratory, having a measured muon flux of $\rm 328\pm 10~muons/m^2/day$~\cite{cos_muons}, with installation completed in March of 2017.

We report here on the development and design of the array and initial signal analysis and performance results.

\section{Detector Design and Development}
\label{sec:Devel}

The array is constructed of fourteen p-type coaxial HPGe elements with 70\% relative efficiency each.  The elements are attached to two opposing copper cryostat bases, with seven detectors on each base.  Each base approximates a cylinder with a diameter of about 35~cm, and each is cooled through a copper cold finger, extending through the detector shielding and connected to its own liquid nitrogen (LN2) dewar.   The detector elements are surrounded by individual copper cylinders extending from the cryostat bases, and are arranged in a hexagonal arrangement with the seventh element for each placed in the center.  The elements are individually encapsulated and sealed to the cryostat base with o-ring flanges to allow easy maintenance of individual detector elements without breaking vacuum for the remaining elements.  A single large o-ring seals each of the two main cryostat bases.  The elements within each hexagon arrangement are positioned with approximately 2.5~cm of space between adjacent cryostat housings, allowing placement of sample material between the elements to optimize efficiency.

The entire array assembly is kept in a room with a volume of $\rm 32~m^3$, just large enough for people to access the hardware.  The room has no forced ventilation under normal operation.  We have installed a stand-alone air filter that recirculates air and filters it through a Samsung AX90M air-cleaner utilizing an H13 class HEPA filter operated at maximum flow rate, specified as $\rm 680~m^3/h$.  Using a particle counter located on the opposite end of the room, we have observed class 1000 level of particulates with this filter operating, even while people are working in the room.

\begin{figure}[!htbp]
  \centering
  \includegraphics[width=0.47\textwidth]{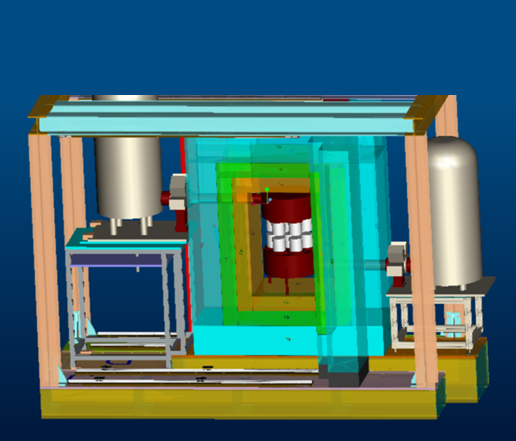}
  \includegraphics[width=0.47\textwidth]{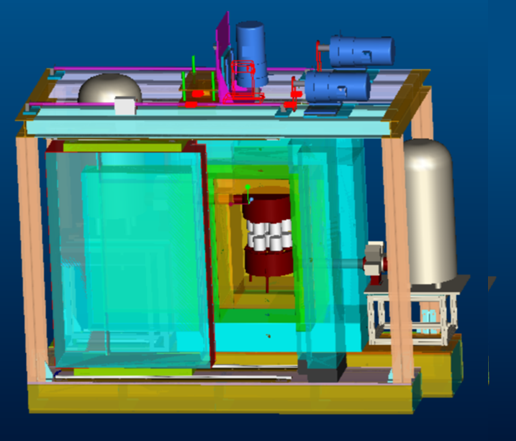}
  \caption{Diagrams of the entire array construction, including shielding and LN2 dewars, without movable doors~(left), and with movable doors~(right). The shielding layers are visible around the array, from inside out as copper, ancient lead, and Goslar lead.  Cold fingers extend from the upper and lower array bases to the dewars.  The shielding doors on the front and back are rolled open and closed on tracks using motors mounted on the top frame. The upper array can be adjusted along with the cold finger, upper LN2 dewar, and a moveable section of shielding.}
  \label{fig:assembly}
\end{figure}

\begin{figure}[!htbp]
  \centering
  \includegraphics[height=7cm]{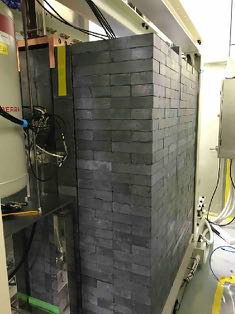}
  \includegraphics[height=7cm]{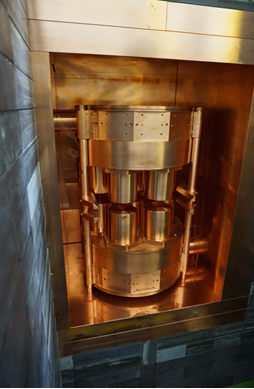}
  \caption{The outer array shielding (left) and the array in the inner copper shielding (right).  The top array LN2 dewar is seen facing the center section of lead shielding with lifting apparatus attached from the top.  The wall of lead toward the left of the photo is one of the two movable side ''doors'' which roll along a track on the floor. The door is rolled open in the view of the array on the right.}
  \label{fig:photos}
\end{figure}

The two array halves are positioned with the coplanar end-caps of the lower half facing upwards and towards the downward-facing end-caps of the upper half, creating a space in between for samples, with about 30~cm in useful diameter and a vertical separation which can be adjusted from 2.5~cm up to 7.5~cm.  A drawing of the entire assembly is shown in Fig.~\ref{fig:assembly} and photos of the completed structure are shown in Fig.~\ref{fig:photos}  Since the detector housings cannot support significant weight, two shelves are mounted to the vertical alignment posts and can be used for supporting samples.  Alternatively sample arrangements can be constructed with supports from the lower cryostat base or the shielding floor.

\begin{figure*}[htbp]
  \centering
 \includegraphics[width=0.6\textwidth]{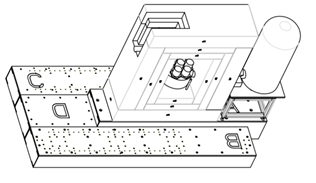}
 \includegraphics[width=0.35\textwidth]{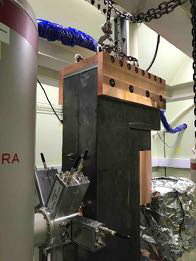}
  \caption{ The diagram (left) shows a conceptual design of the array with the central shielding. Movable shielding doors on the sides are not shown.  The top array and its cryostat are not shown in the diagram.  The shielding penetration for the top array's cold-finger is visible in the upper-left of the shielding.  The vertical shielding column allows the cold-finger and shielding inserted above it to raise and lower with adjustments to the top array height.  The photo (right) shows the missing shielding section above and below the upper cold-finger, with a temporary support chain in place, later replaced with a permanent lifting apparatus.}
  \label{fig:inner_shielding}
\end{figure*}

To achieve the adjustable separation used to optimize sample volume and efficiency, the entire upper assembly can be raised along with its cold finger and LN2 dewar.  Since the upper cold finger penetrates the shielding, a column of shielding, including the copper and lead, above the cold finger can be raised, and shims can be inserted under the cold finger to fill in the gap produced when raising it.  Fig.~\ref{fig:inner_shielding} shows a top view of the shielding without the doors, with the movable column removed.  Threaded rods can be used to raise the top array and dewar together, eliminating stress on the cold finger.  The dewars are mounted permanently on adjustable supports, while a purpose-built adjusting tool is mounted between the two array halves during lifting or lowering and is replaced with fixed-length supports after achieving the desired height. Since the detector is targeted at high-sensitivity measurements with long-duration of counting, it is not expected that this operation will be performed very frequently.  

In the final operation, the openings to each side of the array (as seen in Fig~\ref{fig:photos}) have been covered with 3M brand Vikuiti film, and a boil-off nitrogen line is inserted into the shielding interior.  The combination serves to flush Rn contamination from the air quickly and effectively.

In an attempt to detect background sources for feedback to the final design, we initially tested a single element of the array in a similar shielding configuration to the final design, and performed MC background simulations.  The resulting observed backgrounds were inconsistent with backgrounds originating from sources entirely outside of the shielding.  In particular, low energy peaks were over-emphasized.  We thus screened several potential prototype components for the detector construction to search for potential reducible sources of backgrounds.  An initial study of this screening, performed by counting with our underground 100\% HPGe detector at the nearby A6 area, was presented in ref~\cite{array_1st,array_screening}.  In particular \Th[228] was found to be a potential background source in aluminum, and O-rings were found to potentially contribute non-negligibly to all of \K[40], \Th[228], and \Ra[226] sub-chain backgrounds.  Table~\ref{tab:orings} lists the results from HPGe counting of a number of candidate o-ring types.

\begin{table*}[htbp]
\tabcolsep=0.15cm
	\caption{Radioactivity results from single-HPGe screening of candidate O-rings for the array construction.   In all cases we assume that \Th[228] and \Ra[224] (half-life of 3.6 days) are in equilibrium.  A 20\% systematic error is included to account for uncertainty in the detector efficiencies. Uncertainties are one standard-deviation, and limits are 90\% C.L.}
\renewcommand{\arraystretch}{1.2}
\renewcommand{\tabularxcolumn}[1]{>{\small}m{#1}}
\newcolumntype{A}{ >{\raggedright\arraybackslash}m{3.8cm} } 																	
\newcolumntype{B}{ >{\centering\arraybackslash}X } 																	
\begin{tabularx}{\textwidth}{ABBBBB}																	
\hline																	
Supplier, Matrial, and/or Part Num.	&	\K[40] \newline [mBq/kg]			&	\Ac[228] \newline [mBq/kg]			&	\Th[228]  \newline [mBq/kg]			& \Ra[226] [mBq/kg]			\\
\hline																	
COG Viton, Vi 650, green	&$	480	 \pm 	180	$&$		<	180	$&$	96	 \pm 	25	$&$	540	 \pm 	110	$\\
Marco Rubber FKM "white viton" V1012-154	&$	2\,500	\pm	500	$&$	104	\pm	26	$&$	68	 \pm 	16	$&$	1\,800	\pm	400	$\\
Marco Rubber, FKM 75A, Black, P/N: V1000-154	&$		<	400	$&$		<	29	$&$	79	 \pm 	20	$&$	460	 \pm 	100	$\\
Samwon, FKM 70, P/N: OR VA75 10154 SW 	&$	790	 \pm 	250	$&$		<	230	$&$	170	 \pm 	40	$&$	710	 \pm 	150	$\\
O-ring USA Viton, V75	&$	1\,020	 \pm 	290	$&$		<	84	$&$		<	92	$&$	520	 \pm 	110	$\\
Marco rubber EPDM E1055 Peroxide cured	&$	3\,300	 \pm 	700	$&$	460	 \pm 	100	$&$	430	 \pm 	90	$&$	580	 \pm 	120	$\\
Easterns Seals EPDM, BS154EP70 	&$	2\,500	 \pm 	500	$&$		<	61	$&$	31	 \pm 	10	$&$	31	 \pm 	10	$\\
Marco Rubber EPDM E1000 Peroxide cured batch 1	&$	2\,300	 \pm 	500	$&$		<	42	$&$		<	11	$&$	29	 \pm 	9	$\\
Marco Rubber EPDM E1000 Peroxide cured batch 2	&$	860	 \pm 	190	$&$		<	14	$&$		<	9.4	$&$	16	 \pm 	5	$\\
O-ring USA EPDM	&$	3\,300	 \pm 	700	$&$		<	75	$&$	57	 \pm 	17	$&$	74	 \pm 	20	$\\
Polymax EPDM SKU: BS154E70 	&$	8\,600	 \pm 	1800	$&$	12\,700	 \pm 	2500	$&$	12\,200	 \pm 	2400	$&$	7\,100	 \pm 	1400	$\\
ROW Inc. FEP encapsulated silicone 	&$	4\,400	 \pm 	900	$&$	1\,050	 \pm 	220	$&$	1\,040	 \pm 	210	$&$	2\,500	 \pm 	500	$\\
Marco rubber, AFLAS TFE-P	&$		<	1100	$&$	350	 \pm 	100	$&$	120	 \pm 	30	$&$	570	 \pm 	120	$\\
CANBERRA stock o-ring	&$	1\,170	 \pm 	290	$&$	360	 \pm 	90	$&$	210	 \pm 	40	$&$	1\,370	 \pm 	280	$\\
\hline																	
\end{tabularx}																	

  \label{tab:orings}
\end{table*}

The final O-rings used were peroxide-cured EPDM seals sold by Marco Rubber and made with their E1000 compound.  Two batches of these O-rings were purchased with measurements of each shown in Table~\ref{tab:orings}.
Ultimately the final production design removed aluminum in the element housings in favor of copper while maintaining acceptable efficiencies to low-energy gammas. 

\section{Electronics and Signal Processing}
\label{sec:electronics}

The array elements are biased by three iseg-brand NHS6060P positive high-voltage programmable supplies, each supporting up to six-channels of output.
The supplied CANBERRA pre-amplifiers provide an HV inhibit signal which is triggered when one of the detector elements becomes too warm.  These signals are connected back to the HV inhibits for the corresponding channels.  

\begin{figure}[htbp]
  \centering
  \includegraphics[width=\textwidth]{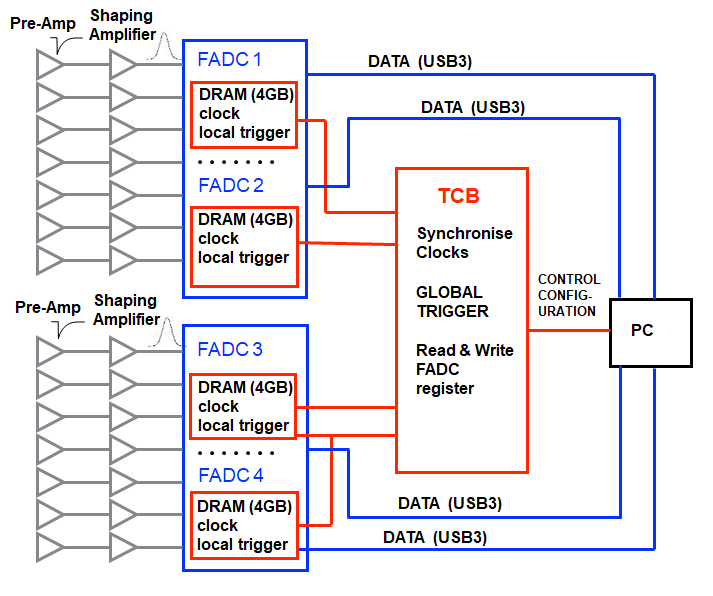}
  \caption{Electronics diagram. Outputs of shaping amplifiers are digitized by four flash ADC modules, all synchronized by a trigger control board that generates global triggers to initiate data transfer to the acquisition computer (PC).}
  \label{fig:electronics}
\end{figure}

The readout electronics are shown in Fig.~\ref{fig:electronics}.  The pre-amp outputs are connected to shaping amplifiers (CANBERRA 2026) with $\rm 6~\mu s$ shaping time. The outputs of the shaping amplifiers are connected to analog inputs of four  12-bit flash analog to digital converters (FADC500, NOTICE) capable of operating at up to $5\times 10^8$ samples per second.  Each module takes four channels of input and two modules are combined within a single physical hardware unit.   A NOTICE supplied trigger control board (TCB) maintains synchronization of the 4~GHz clocks of the FADC modules, and interprets local triggers from each module to generate a global trigger.  A global trigger results in data frames being sent directly from the FADC's to the data-acquisition (DAQ) PC by USB~3.0 connections.

\section{Analysis and Performance}
\label{sec:Analysis}

\begin{figure}[htbp]
  \centering
  \includegraphics[width=\textwidth]{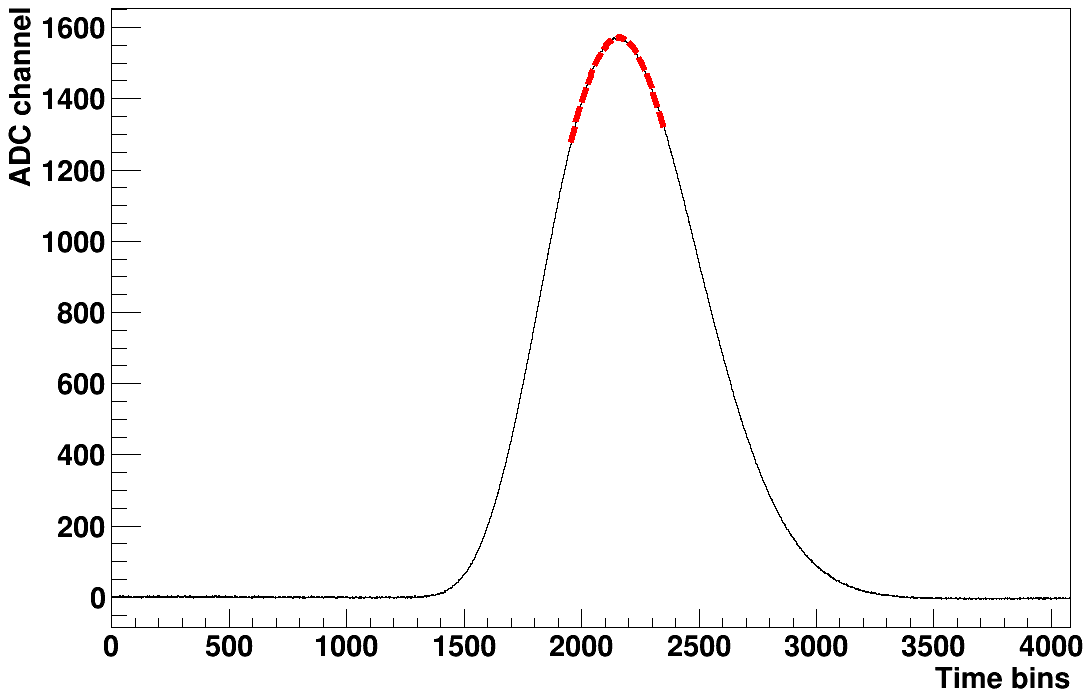}
  \caption{A typical waveform (solid curve) from the shaping amplifier of an array element, as digitized by the FADC500.  A Gaussian fit (dashed curve) is used to determine the pulse height.  Time bins span $\rm 16~ns$ per bin. }
  \label{fig:waveform}
\end{figure}

Waveforms are processed by fitting a Guassian peak to the data in a range within 200 bins before and after the maximum peak height.  The pedestal is sampled and subtracted.  A typical waveform fit is shown in Fig.~\ref{fig:waveform}.  Calibrations are applied separately for each detector element, and a ROOT analysis file is constructed such that each event contains the energy recorded by every detector.  

\begin{figure}[htbp]
  \centering
  \includegraphics[width=\textwidth]{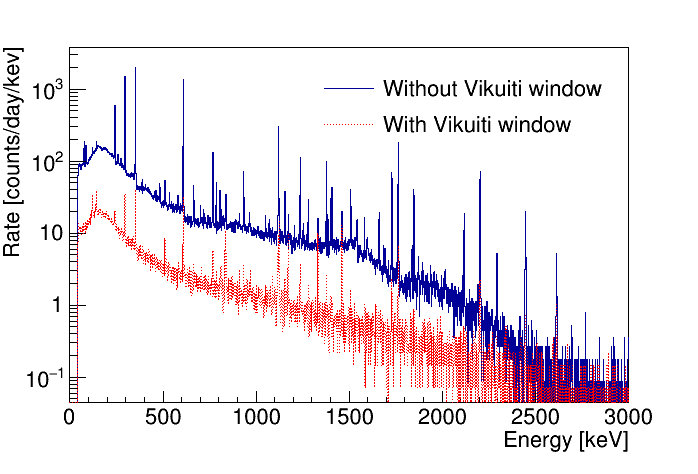}
  \caption{Background spectra for the detector array before (solid) and after (dashed) adding the Vikuiti window over the door area.  The spectra are for the total of all energies recorded in any single event.   }
  \label{fig:bgspec}
\end{figure}

Background data were initially obtained with untreated air surrounding the detector and an LN2 boil-off flush line inserted into the detection chamber for flushing.  Studies of radon and dust levels in the detector room were reported in Ref.~\cite{Mo100_ICHEP}. To investigate effects of radon or dust, as described in Sec.~\ref{sec:Devel}, we covered the detector sides with 3M brand Vikuiti film in an attempt to improve the seal and flushing of the detection chamber.  A second background run was taken in this configuration and indeed resulted in significantly reduced background levels. Both background spectra are shown in Fig.~\ref{fig:bgspec}. Reductions were noticed in both the \Ra[226] and \Th[228] decay-chain activities as shown in Table~\ref{tab:bkgds}.  While \Ra[226] daughters can be transported far in air through the movement of \Rn[222], the shorter half-life of the thorium-related Rn isotope, \Rn[220], creates an expectation that such effects would be reduced. One possible explanation is the reduction of dust achieved both by the application of the Vikuiti film and by the introduction of the  room air filter described in Sec.~\ref{sec:Devel}.  

Detector resolutions were measured in the first month of full operation using the intrinsic 1\,460~keV peak from \K[40].  FWHM values were determined from Gaussian fits and are reported in table~\ref{tab:res}.

\begin{table*}[!htbp]
	\begin{center}
	\caption{Count rates for gamma peaks generated from the \U[238], \Th[232], and \K[40] decay chains.  Results are shown before and after covering the door area with a Vikuiti window, effectively producing a better air seal, while still allowing a nitrogen gas flush.}
	 \begin{tabular}{@{\extracolsep{4pt}}lccc@{}}											
\hline											
	&		&	\multicolumn{2}{c}{Rate [cnts/day]}							\\
\cline{3-4}											
Decay	&	Gamma Energy [keV]	&	No window			&	With Vikuiti window			\\
\hline											
\multirow[t]{1}{*}{\Pb[212]}	&	238.6	&$	36	\pm	3	$&$	5.1	\pm	0.9	$\\
\multirow[t]{3}{*}{\Pb[214]}	&	242.0	&$	511	\pm	5	$&$	9.7	\pm	1.0	$\\
	&	295.2	&$	1\,223	\pm	8	$&$	23.5	\pm	1.2	$\\
	&	351.9	&$	2\,200	\pm	10	$&$	41.7	\pm	1.4	$\\
\multirow[t]{4}{*}{\Bi[214]}	&	609.3	&$	1\,412	\pm	8	$&$	34.1	\pm	1.2	$\\
	&	1\,120.3	&$	354	\pm	4	$&$	9.8	\pm	0.7	$\\
	&	1\,764.5	&$	414	\pm	4	$&$	8.7	\pm	0.6	$\\
	&	2\,204.8	&$	113.4	\pm	2.3	$&$	3.1	\pm	0.4	$\\
\K[40]	&	1\,460.8	&$	19.3	\pm	1.2	$&$	17.1	\pm	0.8	$\\
\multirow[t]{2}{*}{\Ac[228]}	&	911.2	&$	6.2	\pm	1.1	$&$	1.7	\pm	0.4	$\\
	&	969.0	&$	3.6	\pm	1.0	$&$	1.6	\pm	0.4	$\\
\Pb[212]	&	238.6	&$	35.9	\pm	2.7	$&$	5.1	\pm	0.9	$\\
\multirow[t]{2}{*}{\Tl[208]}	&	583.2	&$	12.5	\pm	1.5	$&$	1.0	\pm	0.4	$\\
	&	2\,614.5	&$	8.7	\pm	0.6	$&$	1.55	\pm	0.25	$\\
\hline											
\end{tabular}											

	\end{center}
  \label{tab:bkgds}
\end{table*}

\begin{table}[htbp]
	\begin{center}
	\caption{FWHM resolutions of all detector elements, obtained from Gaussian fits to the 1\,460~keV peak of \K[40] decay.   The detectors are numbers as 1 -- 7 for the bottom array and 8 -- 14 for the top array.  The values were obtained in the first month of operation.}
	 	\begin{tabular}{@{\extracolsep{4pt}}cccc@{}}
	\hline
	\multicolumn{2}{c}{Bottom}			&	\multicolumn{2}{c}{Top}			\\
	\cline {1-2}\cline{3-4}							
\multicolumn{1}{c}{		}&\multicolumn{1}{c}{	FWHM Resolution	}&\multicolumn{1}{c}{		}&\multicolumn{1}{c}{	FWHM Resolution	}\\
\multicolumn{1}{c}{Detector}&\multicolumn{1}{c}{	[keV]	}&\multicolumn{1}{c}{	Detector}	&\multicolumn{1}{c}{	[keV]	}\\
	\hline							
	1	&	1.99	&	8	&	1.93	\\
	2	&	2.01	&	9	&	2.04	\\
	3	&	1.89	&	10	&	2.00	\\
	4	&	1.93	&	11	&	2.07	\\
	5	&	2.05	&	12	&	1.99	\\
	6	&	1.96	&	13	&	1.91	\\
	7	&	1.91	&	14	&	2.10	\\
	\hline
	\end{tabular}							

	\end{center}
  \label{tab:res}
\end{table}

\section{Conclusion}
\label{sec:conclusion}

An array of HPGe detectors with a total efficiency much higher than a standard single-element 100\% HPGe detector, and with competitive backgrounds, was developed and installed at the underground lab in Yangyang, Korea.  The high solid-angle of coverage also results in high probability for detection of coincident gammas for sources placed in the detector interior, adding a powerful background discrimination tool.  The detector resolution and backgrounds measured are well suited to extended low-background counting.  Sensitivities to activities in sample materials are very dependent on the sample details and the optimization of each sample geometry.   Details of coincidence analysis and resulting sensitivities for the first measurements will be reported separately.  With optimized counting with the array, we can expect to achieve sensitivities to \Th[228] and \Ra[226] several times lower than those achieved by typical screening with our standard 100\% detectors.  Furthermore the high coincidence efficiency provides an expectation that we can search for physics at previously unachieved sensitivity levels, for example, searches for decays of \Ta[180m]~\cite{Ta_sim}.
\clearpage

\section*{Acknowledgements} 
This work was supported by the Institute for Basic Science (IBS) funded by the Ministry of Science and ICT, Korea (Grant id: IBS-R016-D1).  We would like to thank Pascal Quirin and his team at CANBERRA, now MIRION Technologies, for their extensive help and cooperation throughout the design process and installation.

\bibliographystyle{elsarticle-num}
\bibliography{hpgearray}

\end{document}